\documentclass[
    sigconf,
    anonymous=false,  
    nonacm=true,      
    timestamp=false,  
    review=false,     
]{acmart}

\AtBeginDocument{%
    \providecommand\BibTeX{{%
        \normalfont B\kern-0.5em{\scshape i\kern-0.25em b}\kern-0.8em\TeX}}}

\usepackage{enumerate}
\usepackage{subcaption}
\usepackage{xcolor}

\newcommand{\smallheadline}[1]{\vskip 3pt \noindent {\textbf{#1}}}

\begin{document}

\title{Flood \& Loot: A Systemic Attack On The Lightning Network}

\author{Jona Harris}
\affiliation{%
  \institution{The Hebrew University of Jerusalem}}
\email{jonahar@cs.huji.ac.il}

\author{Aviv Zohar}
\affiliation{%
  \institution{The Hebrew University of Jerusalem}}
\email{avivz@cs.huji.ac.il}

\begin{abstract}
    The Lightning Network promises to alleviate Bitcoin's known scalability
problems. The operation of such second layer approaches relies on the
ability of participants to turn to the blockchain to claim funds at any
time, which is assumed to happen rarely.

One of the risks that was identified early on is that of a wide systemic
attack on the protocol, in which an attacker triggers the closure of many
Lightning channels at once. The resulting high volume of transactions in the
blockchain will not allow for the proper settlement of all debts, and
attackers may get away with stealing some funds.

This paper explores the details of such an attack and evaluates its cost and
overall impact on Bitcoin and the Lightning Network. Specifically, we show
that an attacker is able to simultaneously cause victim nodes to overload
the Bitcoin blockchain with requests and to steal funds that were locked in
channels.

We go on to examine the interaction of Lightning nodes with the fee
estimation mechanism and show that the attacker can continuously lower the
fee of transactions that will later be used by the victim in its attempts to
recover funds - eventually reaching a state in which only low fractions of
the block are available for lightning transactions. Our attack is made
easier even further as the Lightning protocol allows the attacker to
increase the fee offered by his own transactions.

We continue to empirically show that the vast majority of nodes agree to
channel opening requests from unknown sources and are therefore susceptible
to this attack.

We highlight differences between various implementations of the Lightning
Network protocol and review the susceptibility of each one to the attack.
Finally, we propose mitigation strategies to lower the systemic attack risk
of the network.

\end{abstract}



\maketitle

\section{Introduction}\label{sec:introduction}
The Lightning Network~\cite{poon2016bitcoin} and other second layer
solutions~\cite{decker2015fast,lind2016teechan,green2017bolt,heilman2017tumblebit} 
have been suggested as a solution to Bitcoin's long-known scalability 
issues~\cite{decker2013information,croman2016scaling,zohar2017securing,
bagaria2018deconstructing,sompolinsky2015secure}. The Lightning Network promises
to increase both the number of transactions that can be processed, and the 
latency per transaction. In contrast, Bitcoin's transaction processing is 
bounded by its block creation rate, and its block size limit, which is currently 
1MB (although a slightly more nuanced and widely adopted consensus rule has
been introduced in SegWit~\cite{bip-141-segwit}).\footnote{SegWit defines a new 
concept of block \textit{weight} and redefines the network rule to a maximum 
block weight of 4M units}

The Lightning Network, along with other payment channel networks, aims to move
the majority of transactions off-chain, with the guarantee that the state of 
channels can be committed back to the blockchain at any time, in a trustless 
fashion. It thus solves the problem of a limited transaction rate by limiting 
the communication on payments to a much smaller set of participants, and lowers 
the number of interactions with the blockchain. 

A major assumption that underlies the trustless operation of the Lightning 
Network is that participants will be able to post transactions to the blockchain 
if they are faced with malicious or non-responding peers. A concern that was 
raised long-ago is that under conditions of blockchain congestion, this may not 
be the case: many transactions that are meant to secure the funds of 
participants will not be admitted promptly to the blockchain, which will allow 
attackers to steal money. 

In this paper we explore and analyze this attack scenario. We lay concrete steps 
for executing a successful large-scale attack that both creates the congestion 
effect and exploits it to steal funds from victims. The reason that the attacker 
is guaranteed to steal funds with a large-scale attack is that victims have a 
limited amount of time (measured in blocks) to execute channel closures 
successfully before the attacker succeeds in claiming their money for himself.

We show that the attack does not need to be as widespread as it may initially 
appear. It can be executed on relatively few channels, without any significant 
costs to the attacker. Specifically, we show that even if the maximal number of
victims' transactions get into blocks (up to the block's limit), attacking
only 85 channels simultaneously \emph{guarantees} the attacker will be able to 
steal some funds, and that each additional attacked channel will have its funds 
stolen as well. 

In fact, it is rarely the case that Lightning participants pay sufficiently high 
fees to fill up entire blocks. We explore the fee determination mechanism and 
show that effectively a much smaller number of simultaneously attacked channels
is required for the attacker to succeed. 

Our attack technique leverages the way multi-hop payments are executed in the 
network to trigger several conditions that combine to assist the attacker:
\begin{itemize}
    \item We manage to greatly inflate the number of transactions needed to 
    successfully close a channel by utilizing many multi-hop payments through 
    the channel.
    \item Honest nodes need to claim funds within a 
    limited number of blocks, otherwise they become available to an attacker. The parameter ruling this is not set too high, otherwise funds 
    are locked for long periods of time in other cases. The various implementations use different parameters for this timeout. For example, LND uses a value of 40 blocks.
    \item All implementations do not make use of the full amount of time available to them to deploy closure transactions, making the attack substantially easier. For example, LND initiates the closure of channels only 10 blocks before the expiration time. 
    \item The timeout for retrieving funds is known in advance and thus attacks 
    can be timed so all victims attempt to send transactions to the blockchain 
    at the same time.
    \item Fees offered by these transactions are pre-set and the timing for the 
    attack can be chosen by the attacker (who will choose to attack at times 
    when the blockchain is congested and high fees are required for acceptance).
    \item Unlike the victims, the attacker \emph{can} increase the fee on his 
    transactions, and replace any remaining transactions made by the victims to 
    claim their funds (after the timeout).
\end{itemize}

We now provide a brief description of the attack that we evaluated. A more 
detailed description appears in Section~\ref{sec:attack}.

\smallheadline{Outline of the attack:}

\smallheadline{I. Setup and Wait}
    The attacker creates channels from a node we designate as his
    ``source node'' to many victim nodes, and locks funds into these channels.
    This step is preferably done when blockchain fees are low, both to save 
    costs and to set the channel's feerate to a low value.
    The attacker then prepares a node on the Lightning Network that is able to
    receive funds (i.e., making sure the other side of its channels has
    sufficient liquidity). We designate this node as the ``target node''. This
    can be achieved by opening channels with other nodes (not necessarily 
    victims) and sending money out to purchase goods or to deposit at exchanges.
    The attacker waits for an opportune moment to attack, predicting that
    blockchain congestion conditions will last for several dozens of blocks.

\smallheadline{II. Initiating Payments}
    The attack is then launched by sending multiple lightning HTLC payments 
    from the source node to the target node using as much liquidity as possible 
    and spreading it out over as many HTLC payments as it is allowed to use. 

\smallheadline{III. Accepting Payments}
    Once all HTLCs have reached the target node, it accepts the payments and
    responds by sending back the secrets required to guarantee receipt of the
    funds. As a result, the target node's channel is left without open HTLCs.
    HTLC secrets make their way back to the victims that attempt to send them
    back to the source node. The source node does not respond, leaving the
    channels with open HTLCs that the victims can only claim on the blockchain.

\smallheadline{IV. Collect Expired HTLCs} 
    As the HTLCs' timeout approaches, the victims attempt to close the
    channels with the source node and claim all HTLCs on the blockchain. This
    means they publish many blockchain transactions all at once. Some of these
    will fail to enter into blocks by the timeout due to congestion and high
    blockchain fees. The attacker claims any HTLCs that remain past expiration,
    using the replace-by-fee policy, and raises the fee of his own transactions
    to be higher than that of the victims' transactions.

\begin{figure}
	\centering
	\includegraphics[width=\columnwidth]{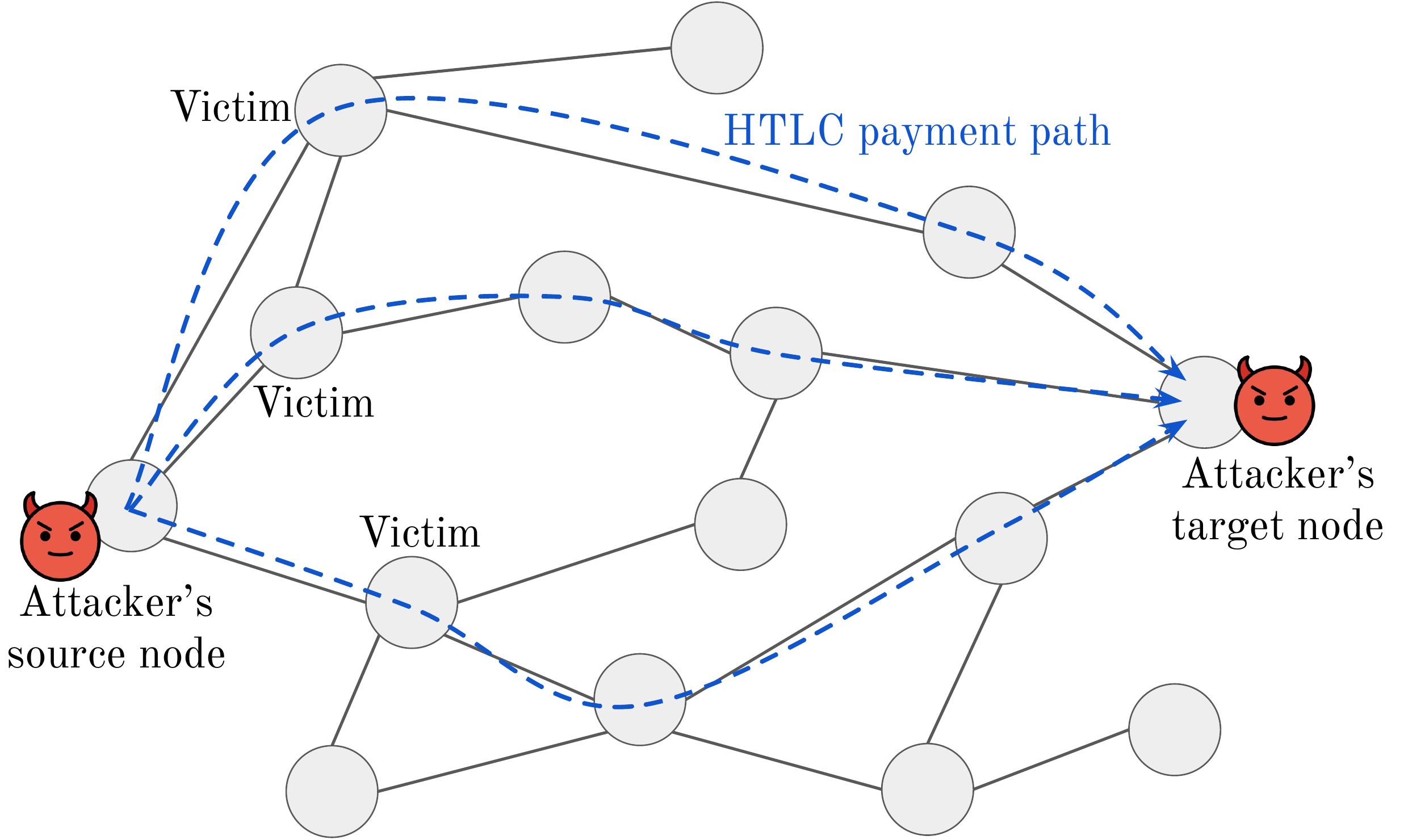}
	\caption{Attacker's and victims' nodes with payment routes from the source
		node to the target}
	\label{fig:general-attack-topology}
\end{figure}

\vskip 10pt
Figure~\ref{fig:general-attack-topology} shows an example topology of Lightning
channels. It emphasizes the paths of the HTLC payments and how victims are
connected with the attacker. We note that whenever the attacker has failed, the
only cost that he suffers is that of the blockchain fees involved with opening
and closing channels with his victims. The funds locked in the channels are
retained and can be reused.

Usually, attackers can reduce the minimal number of victims required for 
executing a successful attack, by exploiting fluctuations in blockchain fees.
We thus go on to explore the feerate estimation mechanism used 
by the different lightning implementations. Using real estimation results 
collected over a period of over two months, we show that the fee used for the 
commitment and HTLC-claiming transactions fluctuates greatly and has significant
influence on the success of the attack.

To demonstrate the steps of the attack outlined above, we report on our 
prototype implementation of an attacker node. Our prototype was evaluated on a 
locally established lightning network, running over a Bitcoin regnet, and
demonstrates that we can exploit the actual clients running today's Lightning 
Network.

A Lightning node can become a victim in the attack if it agrees to open a
channel jointly with the attacker.
We go on to check which nodes in the Lightning Network are susceptible to be
attacked by agreeing to open channels with us.
We report on an experiment that attempted to connect to nodes on the network
and show that a large majority (95\%) of \emph{responding} nodes agree to form such
connections (other nodes refuse for often benign reasons, such as
in-progress synching with the blockchain).
We conclude that it is therefore easy
for the attacker to find enough victims to perform a wide-scale attack.
Since nodes are able to open more than one channel with each peer, attackers are
able to steal more funds from each such victim.

A side-effect of the attack is that the blockchain is flooded with transactions
made by lightning nodes, whose fees are not covered by the attacker, but rather
by his victims. This can be leveraged as a tool in other attacks as well (e.g.,
preventing trade and other on-chain activity).

Our final contribution in this paper is proposing several mitigation techniques
that can greatly reduce the chance and scope of a successful attack.

\paragraph{Ethical Concerns}
The attack that we discuss here can be used to steal funds from users on the
Lightning Network. While our specific technique for attacking is detailed here,
the general attack vector was known ahead of time. We note that preprints of
this paper have been shared with the developers of the three main Lightning
implementations before submission.


The code we used for simulating the attack is available on
GitHub\footnote{\url{https://github.com/jonahar/lightning-systemic-attack}},
without the full prototype implementation of the attacker, so as not to provide
attacking tools.

The remainder of this paper is structured as follows:
Section~\ref{sec:background} gives some background on the Lightning Network
and the different transactions that are created as part of operating a payment
channel. 
Section~\ref{sec:attack} describes in detail the execution of the
attack. 
In Section~\ref{sec:evaluation} we analyze the attack, explore how
effective it might be and show the minimal effort required to find potential victims.
In Section~\ref{sec:fees} we study the fees used by nodes
on the Lightning Network (for blockchain transactions) to learn how they affect
the success of an attack. 
Section~\ref{sec:mitigations} discusses several potential ways to
mitigate the attack. 
Section~\ref{sec:related-work} discusses related work, and finally we
conclude this work in Section~\ref{sec:conclusions}.

\section{Background on Lightning Channel Transactions} \label{sec:background}
Each channel on the Lightning Network can be used for multiple transfers.
Each transfer over the channel is executed by exchanging Bitcoin transactions
between the two parties. These transactions, which are generally not sent to the
blockchain are used as a fail-safe, allowing the two participants in the channel
to claim their funds in case their peer misbehaves.

Under normal operating conditions, i.e, when both parties are honest, no more
than 2 transactions should be published throughout a channel's entire lifetime.
Our attack however, takes the victim's channel off of this ``good path'' and
triggers the publication of possibly hundreds of Bitcoin transactions.
In this section we provide background explaining which transactions are created
and used to support a channel's operation and outlining the conditions under
which they are transmitted to the blockchain.

\begin{figure*}
    \centering
    \includegraphics[width=\textwidth,height=0.4\textheight,keepaspectratio]
    {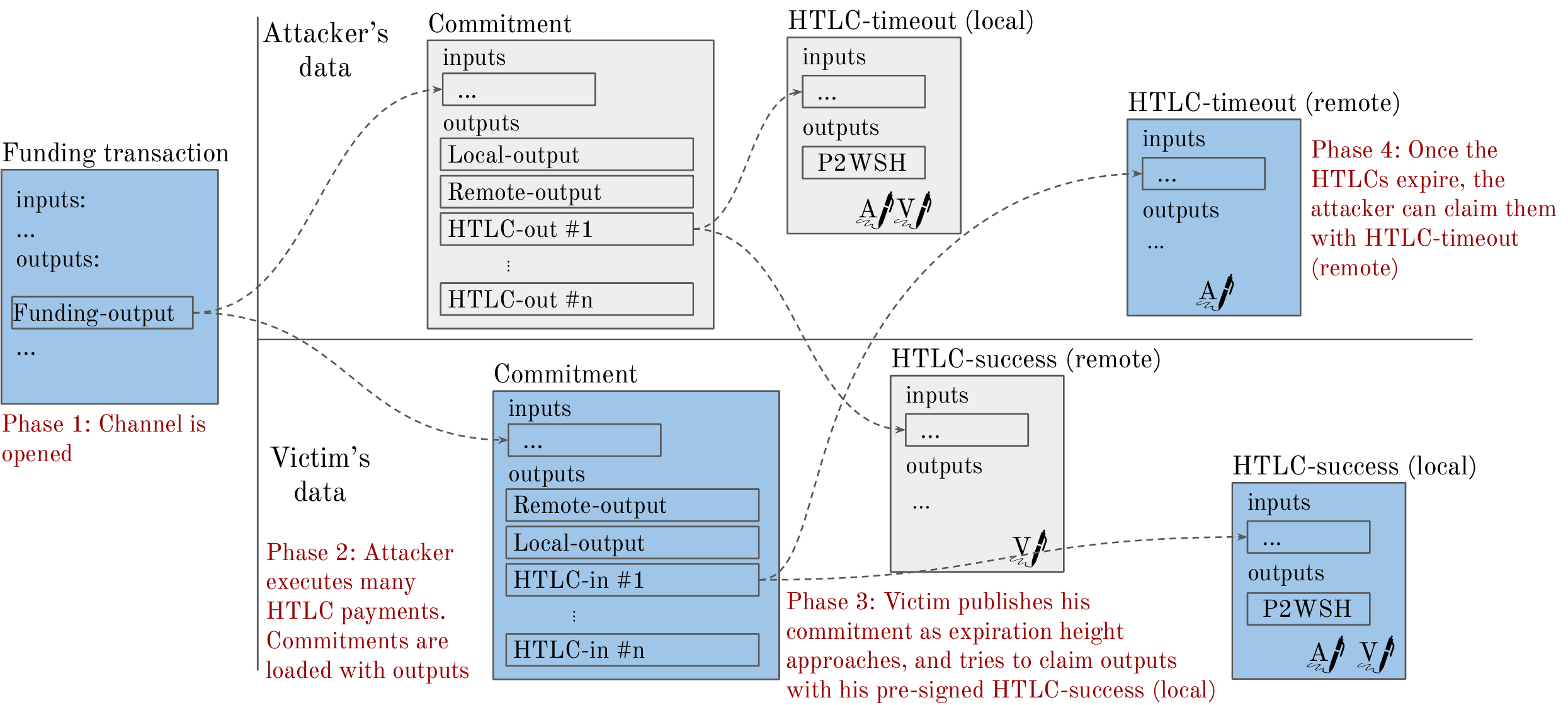}
    \caption{The transactions held by the channel's parties with transactions
    published in the attack marked}
    \label{fig:attack-transactions}
\end{figure*}

\subsection{Channel Establishment}\label{subsec:channel-establishment}
When two parties, Alice and Bob, wish to open a new lightning channel, they
first need to lock some coins to fund it. This is done by creating and
publishing a \textit{funding transaction}. A funding transaction has one
special output (possibly among other outputs) called the \textit{funding
output}, which sends funds to a 2-of-2 multisig script, corresponding to the
private keys of the two parties. This means the funding output can only be
spent by a transaction signed by both Alice and Bob. The value of the funding
output determines the total liquidity in the channel. To guarantee funds can be
extracted from the channel in case one side becomes uncooperative, another
transaction called \textit{commitment} is created and signed by both parties.
A commitment spends the funding output and splits its value between Alice and
Bob, according to the most recent allocation they agreed on (the commitment will
have two outputs - one for Alice and another for Bob). Each of them can publish
the commitment at any time to close the channel and claim their funds, without
any action taken by the other party. For reasons we explain later, each
side holds a slightly different version of a commitment transaction.

\subsection{Balance Redistribution and Commitment Updates}
Alice and Bob can agree on a new way to split the channel's funds between them 
(e.g. if Alice purchases some goods from Bob).
To do so, Alice and Bob create a new commitment that reflects
the changes -- the values of the outputs in the new commitment will match the
new allocation between them.

Both sides must have an assurance that the other side 
will not publish older commitments, especially if the allocation they represent awards that party more money than the last created commitment. This is achieved by the revocation
mechanism, that penalizes participants who publish old commitments.
In Alice's commitment, the output that is destined for her can in fact 
be spent not only by Alice, but also by anyone else who knows a special
\textit{revocation key}. The revocation key can be constructed from two secrets,
held by Alice and Bob (the secrets are commitment-specific). When Alice and Bob
agree on a new commitment, Alice will provide Bob with her secret, allowing him to spend her output in the old commitment. This removes
any incentive Alice might have to publish the old commitment.
To prevent a race, and give Bob a fair chance of spending Alice's output from
an older commitment, the output is time-locked for spend by Alice. Alice can
only spend her output after the commitment achieves a minimum number of
confirmations (determined by the channel-specific parameter
{\verb|to_self_delay|}), while Bob can spend it immediately.
In a similar way, Bob will provide Alice with a secret, that would let her spend
Bob's output in his old commitment transaction.

\subsection{Multi-Hop Payments With HTLCs}
An additional important property of the Lightning Network is its ability to
route payments between participants that do not share a channel.
This is achieved using Hash Time Locked Contracts (HTLC).
HTLCs are conditional expiring payments, that can be routed through 
an untrusted intermediate node.
HTLCs guarantee that no party is able to leave at any moment and stealing other 
node's funds.

Suppose Alice wish to send funds to Carol via an intermediate node Bob, with
whom they both share a channel. Alice will offer Bob the amount of payment
(in addition to a small fee) if he would send that amount to Carol.
The payment starts by Carol generating a secret $s$ (the \textit{preimage}),
computing its hash value $r=h(s)$ and sending $r$ to Alice.
Alice will then make a payment to Bob, of the amount she would like to send 
Carol plus an extra fee for Bob, subject to Bob revealing some secret $s'$ that
hashes to $r$. The only (feasible) way for Bob to claim these funds, is to make 
a similar conditional payment to Carol - requesting her to reveal such $s'$ in 
order to receive the payment. Since Carol holds $s$, she will reveal it 
to Bob and claim the payment from him. Bob will then reveal $s$ to Alice, to
claim the payment from her.
Each conditional payment (by Alice to Bob and by Bob to Carol) has an 
expiration, after which the payment initiator can claim the offered funds back 
to himself.

\paragraph{How HTLCs affect the commitment}
When a new HTLC payment is offered, the commitment must be updated to reflect
the new state, and to enforce the details of the nodes' agreement. This is done
by creating a special output, corresponding to that HTLC, in the commitment
transaction. The HTLC output is removed from the transaction only once it is
resolved, either by the receiver returning the preimage or stating that he
failed to obtain the secret. Since multiple HTLCs may be pending resolution at a
given point in time, a commitment transaction may contain \emph{many} HTLC
outputs (in addition to its two primary outputs for the two sides), resulting in
a very large commitment transaction.

It is important to note that an HTLC output is mirrored in the commitment
transactions of both sides of the channel (See the
\textit{HTLC-out} and \textit{HTLC-in} outputs in the commitments in
Figure~\ref{fig:attack-transactions}).
To facilitate revocation, these
outputs, that correspond to the same HTLC, are slightly different.

In the context of a specific commitment transaction, we refer to the party
holding this transaction as the \textit{local node}, and the other party as 
the \textit{remote node}.

We now drill deeper into the structure of the HTLC-in and HTLC-out outputs as these are relevant to our attack.

\paragraph{The HTLC-out output}
HTLC-out is a commitment output corresponding to an HTLC payment offered by
the local node (the one holding this commitment) to the remote node.
This output enforces all the rules regarding the HTLC payment agreement.
As such, it can be spent in one of the following ways:
\begin{itemize}
    \item By the remote node using the HTLC preimage\footnote{this output can
    also be claimed by the remote node using a revocation key, but we ignore 
    this for brevity}
    \item By the local node, once the HTLC expired
\end{itemize}
The remote node can spend this output by issuing a transaction referred to as 
\textit{HTLC success (remote)}. This transaction should be signed only by the 
remote node, and can be published at any point in time.

The local node can spend this output by a transaction referred to as 
\textit{HTLC timeout (local)}. This transaction can be published only after the 
HTLC expired, and must be signed by \emph{both} parties.
Since the local node should not depend on the remote node, the HTLC-timeout 
(local) is created and pre-signed by the remote node when the HTLC is created.

\paragraph{The HTLC-in output}
HTLC-in output is very similar to an HTLC-out, only that it represents a payment
offered by the remote node. Conditions for spending this output are similar, and
notably, here as well in order to spend HTLC-in using the preimage, the local
node must use a transaction signed by \emph{both} parties, denoted
\textit{HTLC-success (local)}.
The remote node can spend an HTLC-in after the timeout with
a transaction referred to as \emph{HTLC-timeout (remote)}.
Figure~\ref{fig:attack-transactions} depicts the different transactions held by
each party when an HTLC payment is added to the channel.

We shall later see that the fact that HTLC-success (local) transactions require
the signature of both parties will prevent the victim from raising the fees for
its transactions. In contrast, HTLC-timeout (remote) transactions require only a
signature by one node, which will allow the attacker to change its transactions'
fees at will. 

\subsection{Replace-By-Fee Policy}\label{subsec:RBF}
We now briefly explain the mechanism used in Bitcoin to replace unconfirmed 
mempool transactions that we use in the attack. Replace-By-Fee (RBF) is a policy that allows transactions to 
opt-in for replacement by other conflicting transactions if they offer higher fees. 
Replacing transactions in this manner can be useful for dealing with confirmation delays that were caused by setting fees too low.

As specified in the Replace-by-Fee improvement proposal (BIP-125~\cite{bip-125-rbf}),
a transaction is replaceable (opts-in) either if it directly signals that it is replaceable, or if one of its unconfirmed ancestors is replaceable.

The Lightning Network specifications~\cite{bolts} define some transactions used
by the protocol as replaceable. Specifically, HTLC-success (local) transactions
are replaceable, a fact that we utilize in the attack to replace unconfirmed
victims' transactions and steal funds.

\section{The Flood \& Loot Attack} \label{sec:attack}
In this section we describe our attack in greater detail. During the attack, the
attacker is able to steal funds from victims that agree to open channels with him.
The attacker ensures that many channels, loaded with
multiple unresolved HTLC payments, will all be closed at the same time, leading
to a high volume of transactions simultaneously waiting to be confirmed on the
blockchain. Inevitably, such congestion will result in some of the HTLC-claiming
transactions published by the victims failing to confirm before they timeout.
The attacker can claim any funds associated with HTLC payments that timed out in this manner.

\begin{figure}
    \centering
    \begin{subfigure}{\columnwidth}
        \centering
        \includegraphics[width=\columnwidth]{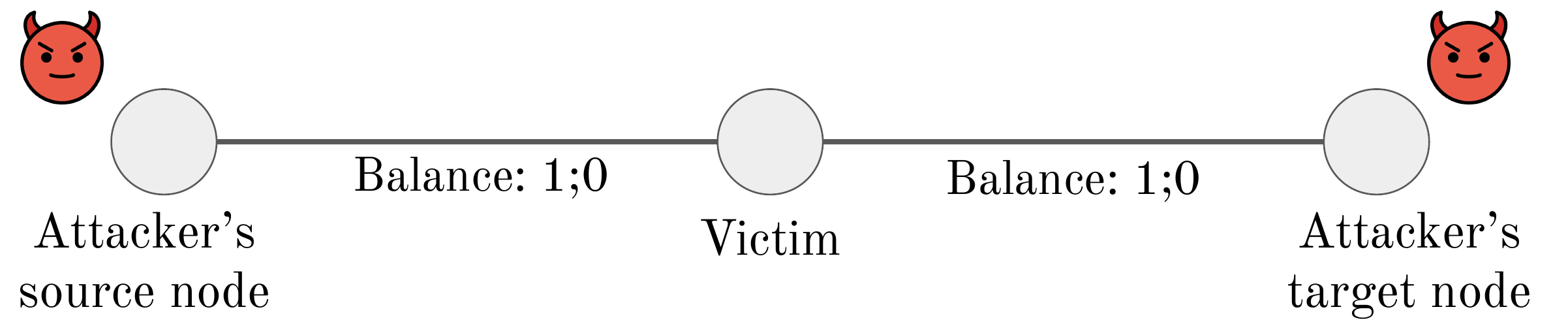}
        \caption{Establishing channels}
        \label{fig:attack-phase-1}
    \end{subfigure}

    \begin{subfigure}{\columnwidth}
        \centering
        \includegraphics[width=\columnwidth]{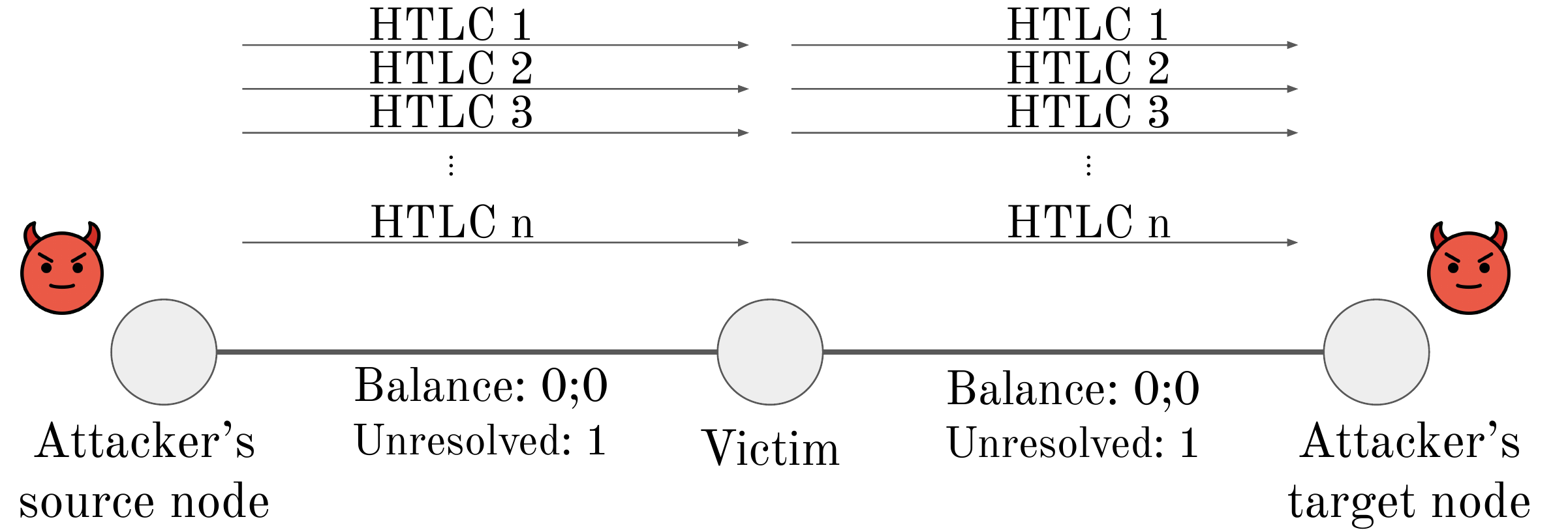}
        \caption{Making HTLC payments}
        \label{fig:attack-phase-2}
    \end{subfigure}

    \begin{subfigure}{\columnwidth}
        \centering
        \includegraphics[width=\columnwidth]{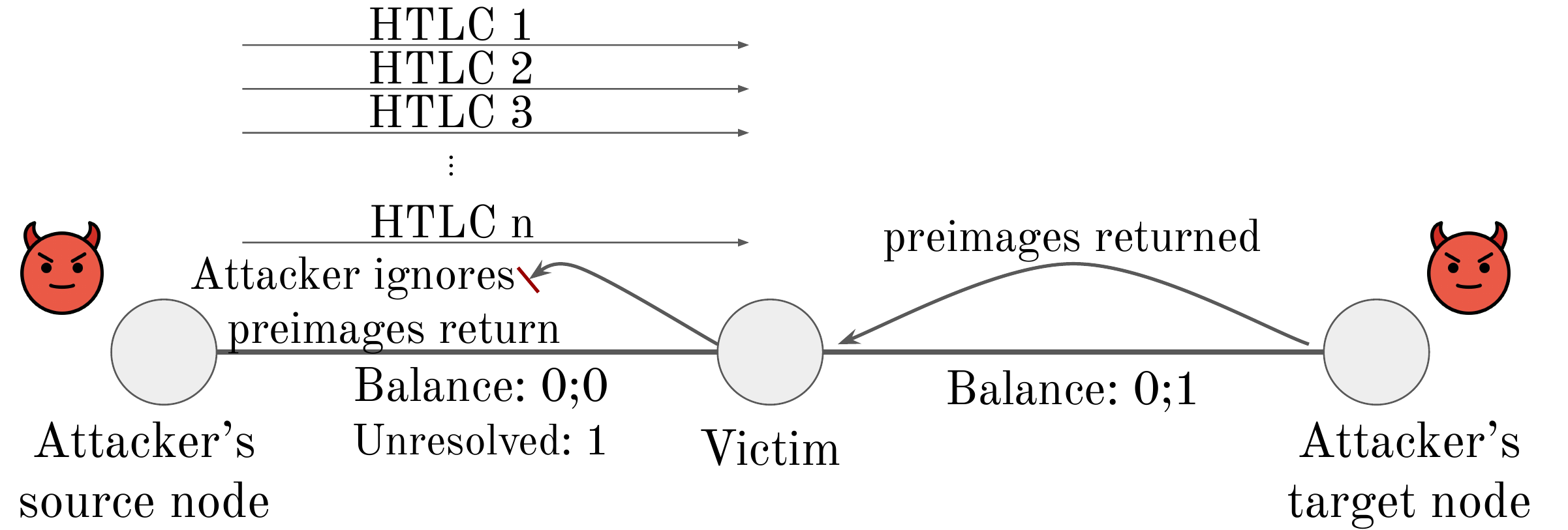}
        \caption{Target node returns the preimages and the source node ignores
        requests to resolve HTLCs}
        \label{fig:attack-phase-3}
    \end{subfigure}

    \begin{subfigure}{\columnwidth}
        \centering
        \includegraphics[width=\columnwidth]{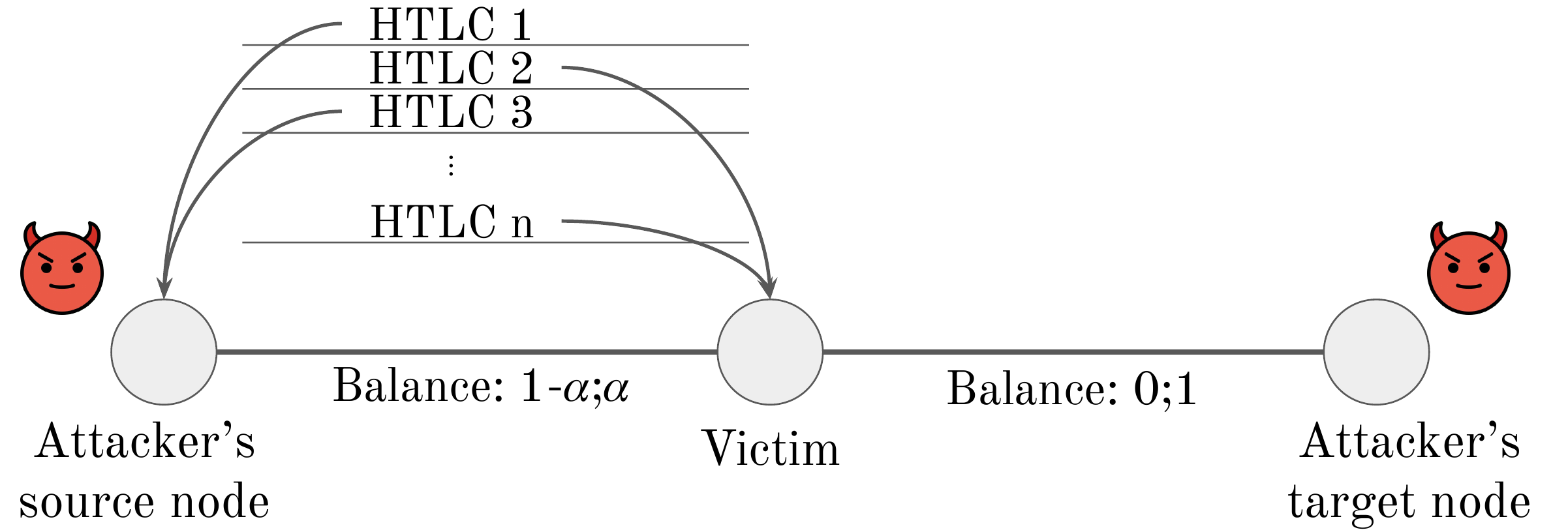}
        \caption{Waiting for expiration to collect HTLCs}
        \label{fig:attack-phase-4}
    \end{subfigure}

    \caption{The main attack phases}
    \label{fig:attack-phases}
\end{figure}

We now detail the four concrete steps the attacker takes from the setup of the
attack until the HTLCs are stolen, as well as the actions of the victims. The
steps are also detailed in Figure~\ref{fig:attack-phases} (with 1 victim for
simplicity).

\smallheadline{Step I: Setup - Establishing Channels}
The attacker controls two lightning nodes, denoted the \textit{source} node and
the \textit{target} node.
The target node should be able to receive a sufficient amount of funds. That
means it has open channels with enough liquidity on its peers' side (this can
be achieved e.g. by buying some goods with a lightning payment or depositing
coins to an exchange that accepts lightning payments).
The source node opens many channels with different victims, possibly even
multiple channels with each victim.
All these channels are initiated and funded by the attacker.

Figure~\ref{fig:attack-phase-1} shows a single victim and the channels involved
in the attack. Note, that the attacker's target node does not have to be
directly connected to the victim (only the source node needs such a direct
channel).

\smallheadline{Step II: Loading Channels With HTLC Payments}
Once all channels are set up, the source node starts sending the maximum
possible number of HTLC payments to the target node, through the victims' channels.

Each channel's side has parameters {\verb|max_htlc_value_in_flight|} and
{\verb|max_accepted_htlcs|}, which are a cap on the total value of outstanding
HTLCs and a limit on the number of HTLCs the other party can offer, respectively.
The attacker chooses the amount for each HTLC payment in a way that
utilizes the maximum funds that can be stolen.
This is done by spreading the maximum transferable amount (excluding fees)
equally across all HTLCs routed through the channel.
Formally, for a channel with balance $b$, a {\verb|max_htlc_value_in_flight|}
value of $M$ and a \ {\verb|max_accepted_htlcs|} value of $a$, the amount of each
payment will be
\begin{equation*}
    htlc\_value =  \frac{min(M, b - fee)}{a}
\end{equation*}
where $fee$ is the fee paid by the commitment transaction.
The exact fee is calculated from the channel's feerate (we discuss the
channel's feerate in more detail in Section~\ref{sec:fees}).

Every HTLC payment offered by the source node is given the same expiration
height. This would result in all attacked channels closing at the same
time.

At that point, the target node does not yet return the preimages to resolve the
HTLC payments, and keeps them all in pending state.
At the end of this stage, all the attacked channels are saturated with pending
HTLCs.

\smallheadline{Step III: Returning Preimages on Last Hop}
After all possible payments were made by the source node, the target node
starts resolving all HTLC payments it received by returning the preimages.
The target node now possesses the total amount of funds sent by the source node and
may close all his channels cooperatively at any time, to use these funds
anywhere else.

Once each victim receives the preimages, it asks the source
node to resolve the HTLCs and move their amount to the victim's side of the
channel. The source node, however, stops any communication with his victims and
refuses to resolve the HTLCs. Any further messages from the victims are ignored.

We note that if the target node does not have enough channels to allow a large
enough number of unresolved incoming HTLCs to accommodate many victims at once,
it could perform steps II and III of the attack one channel at a time, i.e.,
saturate a single channel of the source node with HTLCs, then return all
preimages for that channel and proceed to the next victim channels. The key
point is that preimages of HTLCs that are routed through the same victim
channel must be delayed and then returned together in a single batch.

\smallheadline{Step IV: Collecting Expired HTLCs}
Each attacked channel is now full with unresolved HTLC payments, and the victims
have all the preimages required to claim them. Since the attacker is
uncooperative, they cannot claim these payments through Lightning
interactions, but will have to close their channels, publish their
commitments and claim the HTLCs on the blockchain.

The time at which a victim will publish his commitment to unilaterally close his
channel, is determined by a safety parameter we denote
{\verb|commitment_broadcast_delta|}.\footnote{This parameter was not named
in the specifications, but is described in~\cite{commitment-broadcast-delta}}
The Lightning specifications suggest a value of 7 blocks before the HTLC expires,
but it is an implementation-dependent decision.
Among the 3 major implementations, LND uses the highest value of 10 blocks.
Since a significant majority of nodes on the network run LND, as suggested
in~\cite{lockdown-attack,mizrahi2020congestion}, most victims will publish their
commitments at the same time (height). Along with the commitments, each victim
will also publish many HTLC-success (local) transactions to claim the HTLC
outputs from his commitment, leading to a high volume of transactions trying to
enter the blockchain all at once.

A sufficiently large number of attacked channels guarantees that some of the
victims' transactions will not be confirmed before the HTLCs expire, simply
because of block space constraints.
Once the expiration height was reached, the attacker can also spend any unspent
HTLC output using HTLC-timeout (remote) transactions. These transactions
conflict with any HTLC-success published by the victims, as they try to spend
the same outputs. The attacker is able to replace any victim's transactions
already in mempools, since they are all replaceable using the Replace-By-Fee mechanism, as we explain in
Section~\ref{subsec:RBF}.

Although HTLC outputs can be claimed by the victims even after they expire, the
attacker has a significant advantage at that point.
In order to replace victims' transactions, the attacker must publish his
transactions with higher fees. The attacker knows exactly what fee is used by
his victims (this is derived from the channel's feerate) and can easily create
transactions with the minimal fee required to outcome those of the victims.
The HTLC-timeout used by the attacker should only be signed by him, and he can
set the fee as he wishes. In contrast, the HTLC-success used by the victims must
be signed by \emph{both} parties (and was in fact already signed when the HTLC
was added), and therefore cannot be changed by the victim to pay more fees, without the cooperation of the attacker.

The reason the attacker only needs his signature, while victims need both
their signatures and that of the attacker, is that the published commitments are
those of the victims.
Contrarily, if the attacker's commitment was the one to be published and
confirmed, the attacker would have to use pre-signed transactions, while the
victims could have claimed HTLC outputs without the attacker's signature.
Of course, in this attack, the victims have no choice but to publish their
commitments before expiration, if they want to claim their funds.

Each HTLC output that is successfully claimed by the source node, is stolen from
the recipient of that HTLC, as this HTLC was already forwarded, and claimed, by
the attacker's target node.

\section{Evaluation} \label{sec:evaluation}
To show the feasibility of the attack we simulated it on a local Bitcoin's
regtest network. We implemented a prototype of an attacker's node (a modified 
version of \textit{C-Lightning}) that can stall the return of preimages and ignore 
peers' requests.
We chose \textit{LND} as the client for the victims, as it is the most popular
Lightning implementation used today.\footnote{as suggested
by~\cite{lockdown-attack,mizrahi2020congestion}, more than 90\% of the
nodes on the Lightning Network runs LND}

\begin{figure}
    \centering
    \includegraphics[width=\columnwidth]{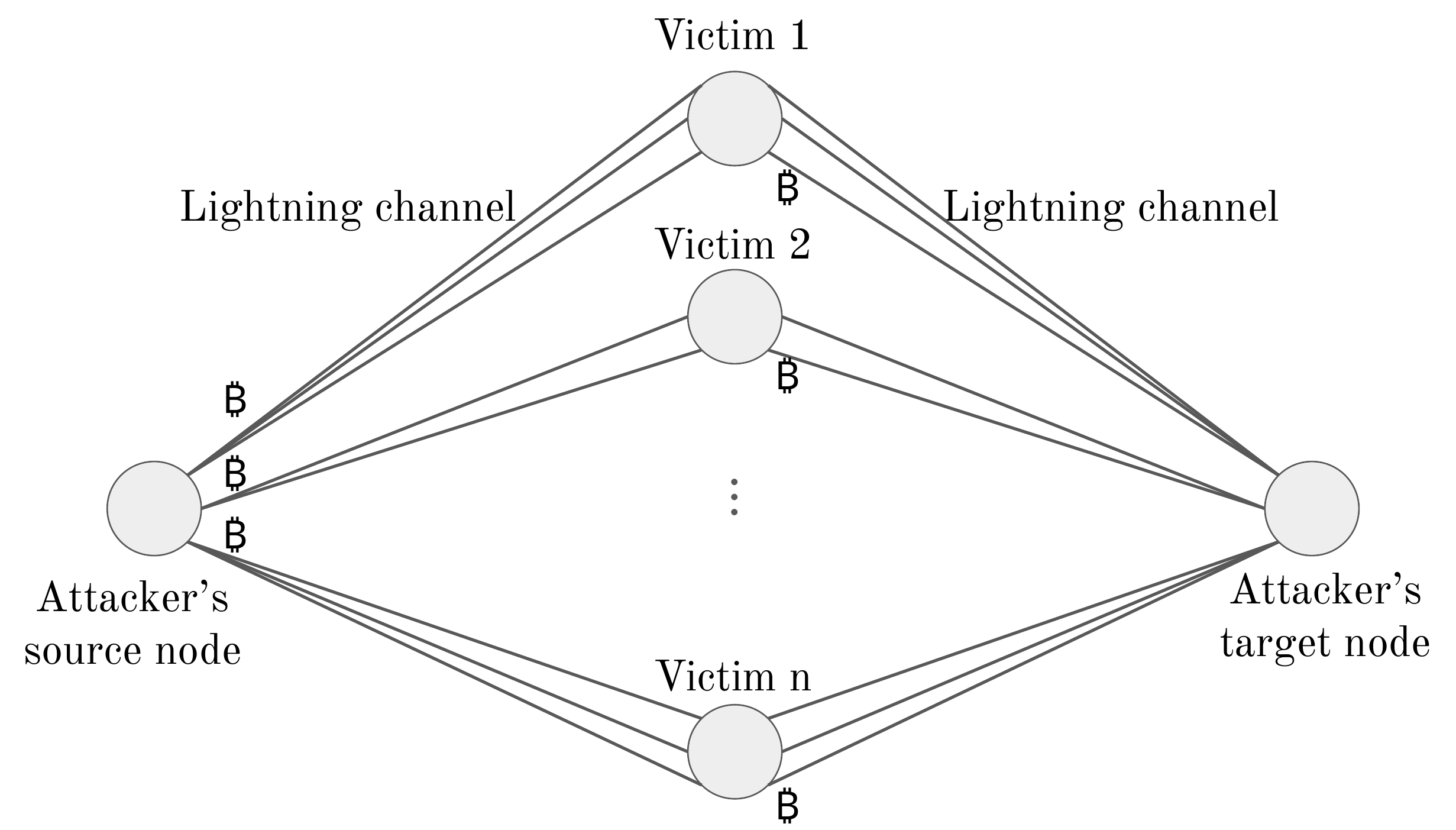}
    \caption{The Lightning channel topology used in our simulations
             with Bitcoin symbols indicating the liquidity allocation within 
             channels}
    \label{fig:simulation-topology}
\end{figure}

Each lightning node in our simulation, both the attacker's and the victims', is
backed by its own full bitcoin node, running \textit{bitcoind}, which are all 
connected to a central bitcoin node responsible for mining. Specifically, the
attacker does not mine blocks and has no way to directly affect which 
transactions are confirmed, other than setting high fees for his own transactions.

One set of channels is established between the attacker's source node
and all victims, and is funded by the attacker. Another set of channels is
established between the attacker's target node and all victims, and is 
funded by the victims, so there is liquidity on their side (in a real world 
attack, an attacker could fund these channels and then move liquidity to the 
other side e.g. by purchasing goods or depositing funds to an exchange with 
lightning payments). The topology used in the simulation is shown in
Figure~\ref{fig:simulation-topology}.

We then route 483 HTLC payments, which is LND's default maximum number of 
unresolved HTLCs at a given moment, from the source node to the target node 
through each channel the source node has. All HTLCs are given the same expiration 
height. Once all payments went through, the target node releases all preimages to 
resolve the HTLCs, gracefully closes his channels and leaves with the total
amount of HTLCs sent to him by the source node (closing the target node's channels
is not required for the attack to succeed, but we do it to show that it can be
easily done and without a significant cost). 
The source node goes into ``silent'' mode and waits until the expiration height to
claim any HTLC outputs that remained unspent.
The victims have exactly 10 blocks for their transactions to be confirmed, from 
the moment they release their commitments, to the expiration, at which the 
attacker starts claiming HTLCs.

\begin{figure}
    \centering
    \includegraphics[width=\columnwidth]{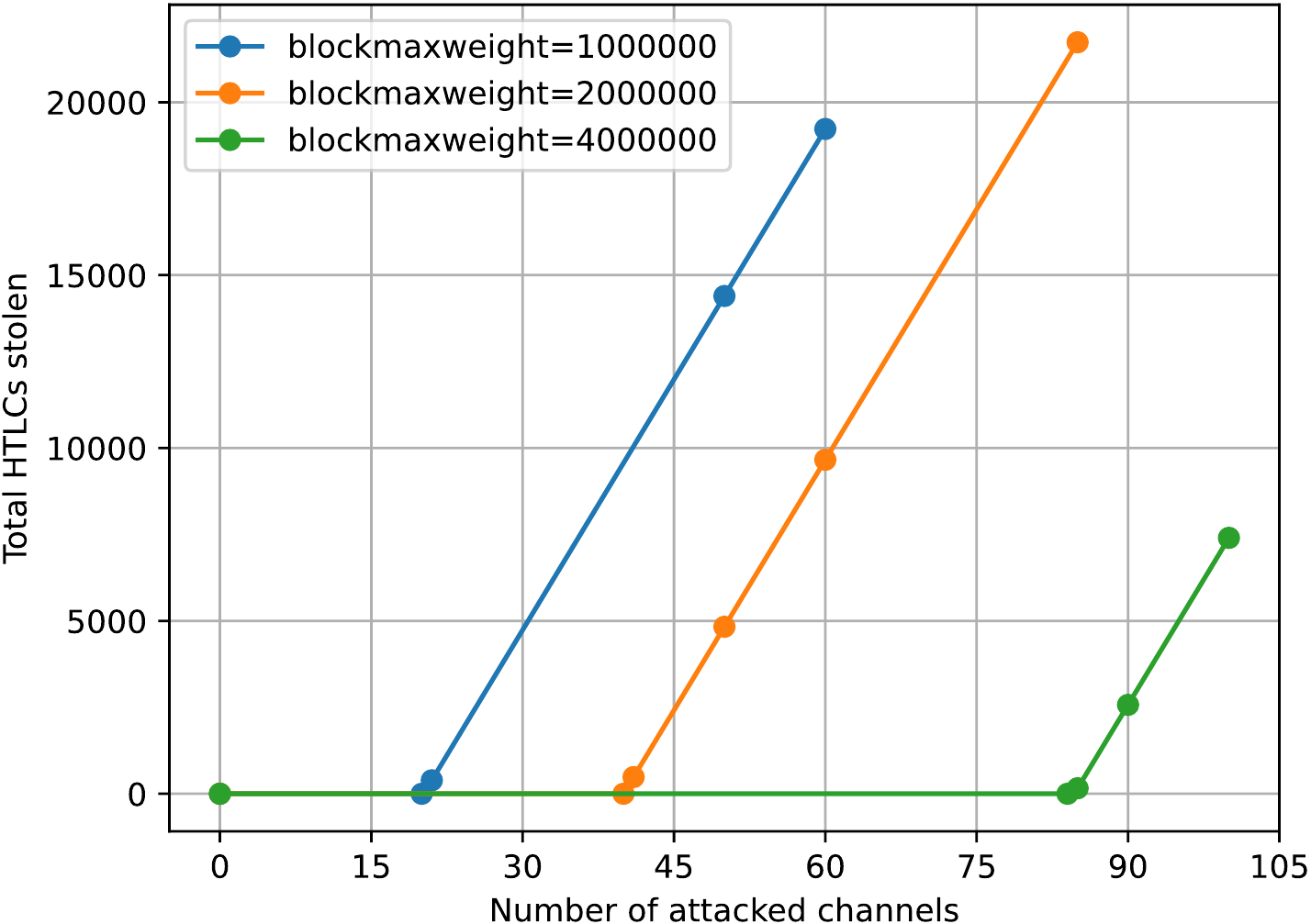}
    \caption{Number of successfully claimed HTLCs vs number of attacked channels}
    \label{fig:num-channels-vs-stolen-htlcs}
\end{figure}

With 100 attacked channels, the attacker was able to successfully steal 7402 
HTLC payments out of 48300 that were made in total. We repeat the simulation
multiple times with different number of attacked channels. The number of 
successfully stolen HTLCs as a function of the number of attacked channels can
be seen in Figure~\ref{fig:num-channels-vs-stolen-htlcs}. We learn that there is
a minimal number of channels, that when passed, the attacker starts to steal
funds, and funds of each additional channel will be stolen as well.
We also conduct more experiments with smaller maximum block weight
(smaller value of {\verb|blockmaxweight|}\footnote{\textit{blockmaxweight} is
the consensus parameter that defines the maximum weight for a valid block})
to represent times of congestion in the network. As we show in
Section~\ref{subsec:feerate-estimations}, there are times in which the effective
block weight available for the victims can be smaller, due to insufficient fees.

\subsection{Finding Potential Victims}
The attack we present requires the attacker to establish channels with multiple 
victims. In this part we show that opening channels with unfamiliar nodes does 
not add a considerable effort to the attacker. Specifically, we show that most 
of the nodes on the network agree to open a channel with us upon our request.

According to the Lightning protocol~\cite{bolts}, a node indicates its acceptance
to open a channel with another by replying with an {\verb|accept_channel|} 
message to an {\verb|open_channel|} request. We try to perform this handshake 
with many nodes on the network.

\begin{figure}
    \centering
    \includegraphics[width=\columnwidth]{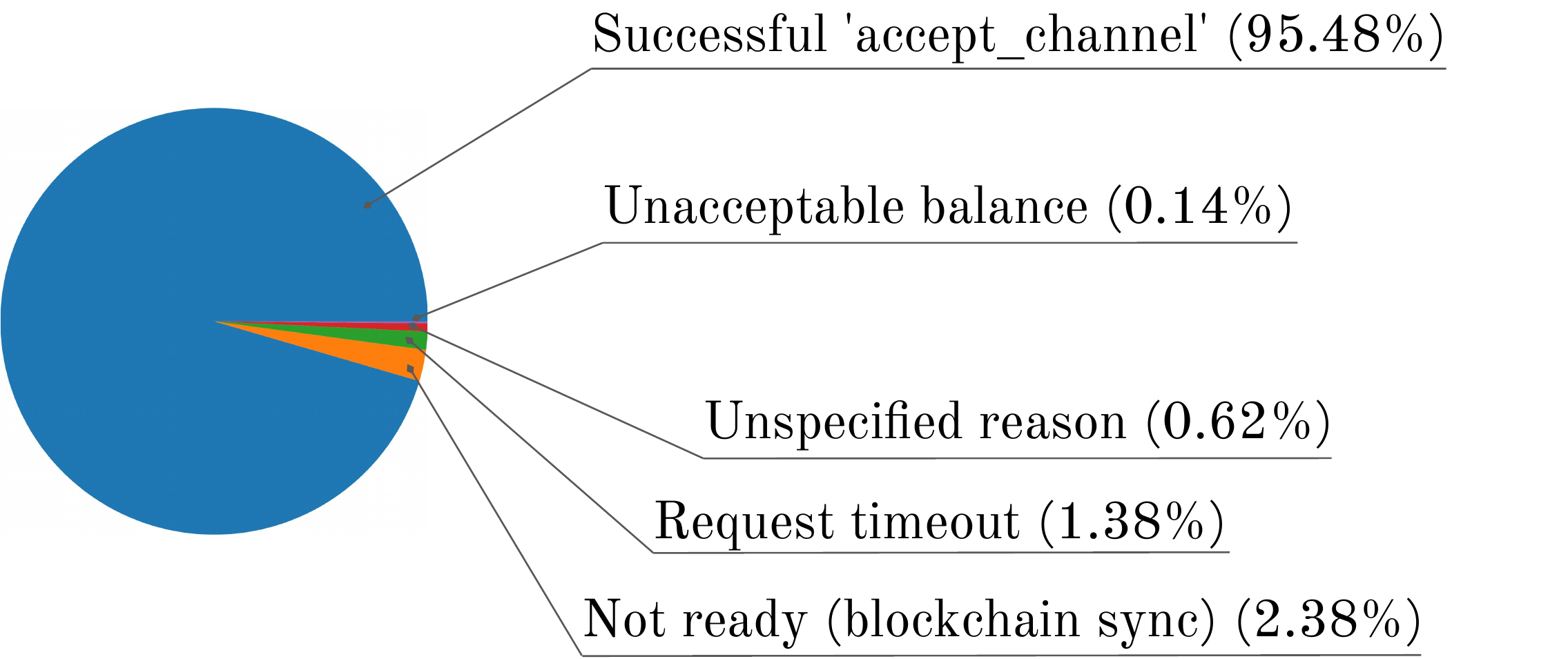}
    \caption{Nodes' responses for open-channel requests}
    \label{fig:handshake-results}
\end{figure}

We start with two lists of the top 50 most connected nodes (nodes with highest
number of open channels) and top 50 nodes with most 
liquidity.\footnote{downloaded from 1ml.com on 2020-06-07}
Excluding duplicates, we remain with
addresses of 64 unique nodes. We connect and send an {\verb|open_channel|}
message to each one of them, to which 63 reply with a successful
{\verb|accept_channel|}.

We move on to find more potential victims. We extract all known IDs and 
addresses from the channels our node heard about and try to connect to all of
them. Out of 4593 nodes, 2102 were responsive and communicating. Most of the
failed attempts to connect were due to timeout or a refused connection 
(indicating that no process was listening at this address, i.e., the node was probably not active).

Out of the 2102 nodes we were able to communicate with, 2007 returned
a successful {\verb|accept_channel|} message, indicating their willingness 
to open a channel. The rest failed for several reasons, as shown in 
Figure~\ref{fig:handshake-results}.

This shows that the vast majority of responsive nodes on the network, 
specifically more than 95\%, are willing to establish a channel by
request, and are therefore susceptible to becoming victims in our attack.

\section{Lightning Channel Fees} \label{sec:fees}
To manage a lightning channel, the two parties must agree on the fee paid 
by each of the transactions that they both sign. Specifically, the fee used for
commitments and HTLC success/timeout transactions. 
When the channel is established for the first time, the parties agree on a
feerate (in Satoshi/KW), denoted the \textit{channel's feerate}, with which they calculate
the exact fee for each transaction. The Lightning
specifications define a deterministic way to calculate the fee for each 
transaction based of the channel's feerate and an estimated weight for the 
transaction (a method for weight estimation is also described in the protocol).
After estimating the transaction weight, the exact fee that should be paid is 
obtained by simply multiplying the feerate and the estimated weight.
The channel's initiator is the one responsible for paying the fees (the fee is 
deducted from his personal output) and is the one proposing the channel's 
feerate when the channel is open.
The other party may reject opening the channel if it thinks the proposed feerate
is unreasonable (too low or unnecessarily large).
Since fees may fluctuate in different times, the lightning protocol specifies a 
way to update the channel's feerate.
The channel's initiator (responsible for paying the fees), and only him, could
propose a new channel feerate by sending an {\verb|update_fee|} message with 
a new proposed feerate, to which the other node must agree.
The protocol states that the other node (the one who did not initiate the 
channel) cannot take any action to update the feerate if it thinks it became
unreasonable, with the rationale that this node is not responsible for paying the fees
(although the node \emph{should} care about the feerate, as it affects the 
node's HTLC transactions).
Any request to update the channel's feerate made by the node who is not the 
channel's initiator, would be considered a deviation from the protocol.

\subsection{Feerate Estimations}\label{subsec:feerate-estimations}
To test how effective the feerates used in lightning channels are, we explore 
the estimation mechanism and see how it affects the confirmation time of
lightning transactions that are published to the blockchain.

\begin{figure}
    \centering
    \includegraphics[width=\columnwidth]{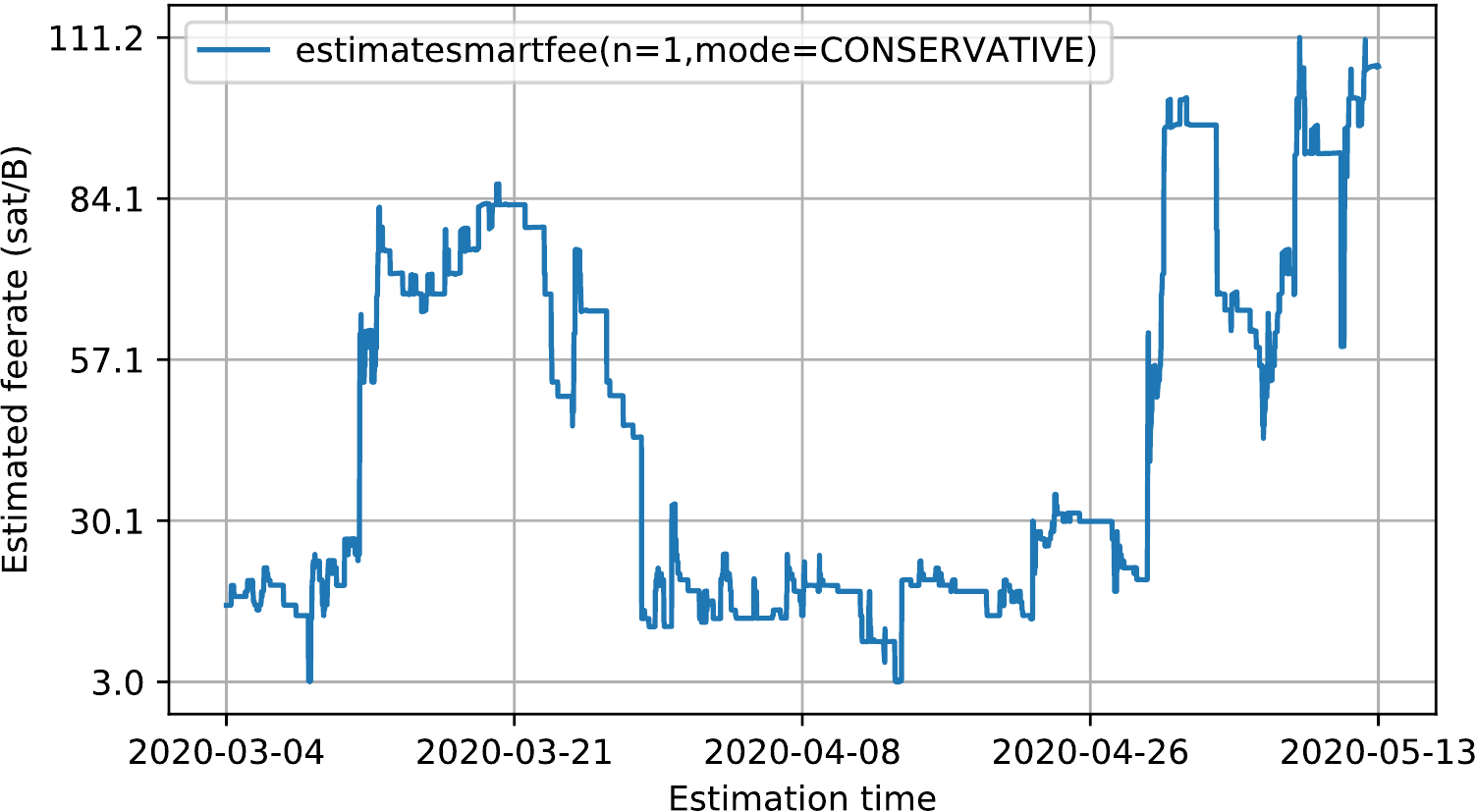}
    \caption{Feerates estimated by bitcoind for immediate confirmation}
    \label{fig:estimated-feerates}

\end{figure}

All 3 major implementations (C-lightning, LND, Eclair) rely on a bitcoin node 
to estimate feerates. This is achieved by querying the 
\textit{estimatesmartfee} method of bitcoind.\footnote{LND also supports the 
\textit{EstimateFee} method of \textit{btcd}}
We continuously queried the estimatesmartfee method of bitcoind over a period of 
time and tried to understand how much block space can be used by
transactions that pay such feerates. In Figure~\ref{fig:estimated-feerates}
we can see the feerate estimations made by bitcoind and how volatile these
estimations could be in a relatively short period of time.

We try to estimate what part of blocks can potentially be claimed by
victims' transactions.
Transactions are normally chosen to enter a block in a decreasing order of their feerate.
We therefore define the available space of block $b$ under feerate $f$ as
\begin{equation*}
    available\_space(f,b) = 4000000 - \sum_{\substack{tx\in b: \\ f_{tx}>f}} weight(tx)
\end{equation*}
Where $f_{tx}$ is the feerate of transaction $tx$.
In practice, some blocks are mined without utilizing their entire available 
weight (4M). The available space in 
such blocks will actually be \emph{smaller} than this estimate, making it overly optimistic, i.e.,  in practice victims will not have
this much available block space.

Next we estimate the average available block space that victims have in the 
limited time window between publishing the commitment and the HTLC expiration.
Since most nodes on the network run LND, we use a window of 10 blocks (LND's
default commitment-broadcast-delta).

\begin{figure}
    \centering
    \includegraphics[width=\columnwidth]
        {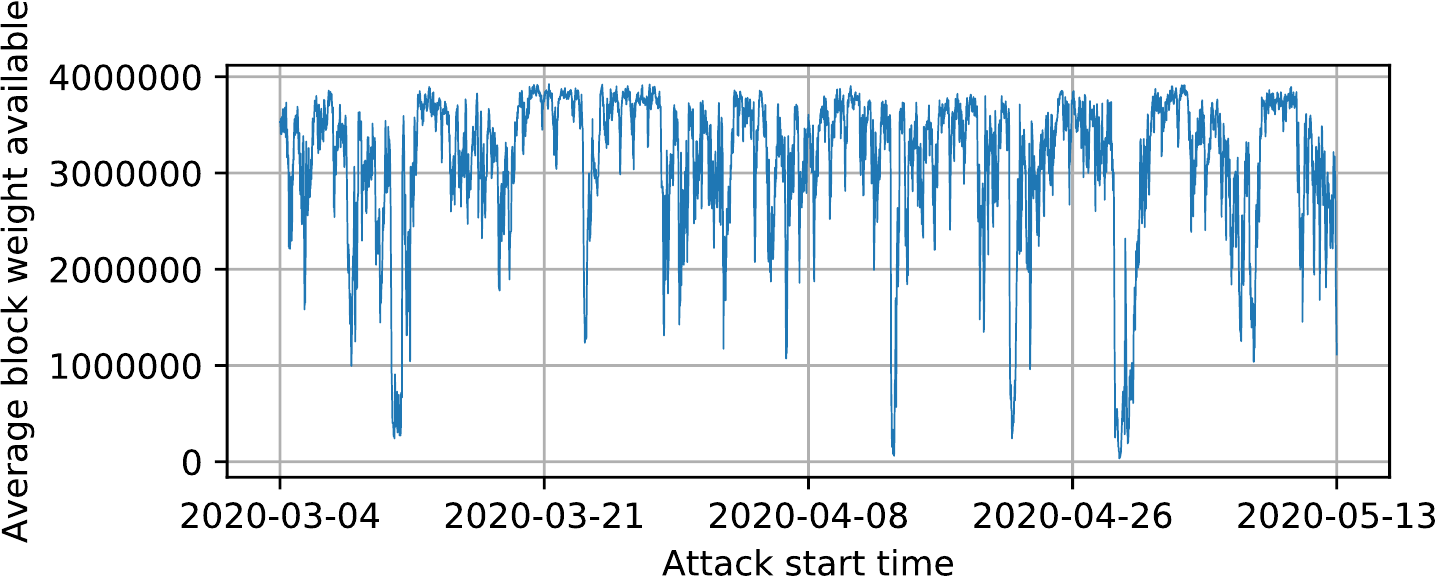}
    \caption{The average block space available for victims on different
    times of an attack}
    \label{fig:attack-start-time-vs-avg-block-space}
\end{figure}

We first look at a simple attack strategy: The attacker opens channels with his
victims and immediately makes all HTLC payments, with an expiration of 100 blocks.
The feerate used on all his channels in this case is the one estimated by 
bitcoind at the moment the channels were opened. In our observed period of over
2 months, this simple strategy shows that victims have an average available block 
space of at least 2M weight units 92\% of the time.
Figure~\ref{fig:attack-start-time-vs-avg-block-space} shows different possible
time points for starting the attack and the average block space available for
victims' transactions, and implies that rarely is it the case that victims can
utilize the entire space of a block.

A less naive attack strategy can further reduce the available block space that
victims will have, by reducing the channel's feerate whenever possible. The
strategy we employ proceeds as follows: The attacker opens channels with his
victims and tries to reduce his channels' feerates for some period of time
before making all HTLC payments, again with an expiration of 100 blocks. As per
the current Lightning protocol, the victims are not allowed to request any
feerate changes, but do agree to any change proposed by the attacker, so long
as the change matches bitcoind's latest fee estimate. Therefore, the attacker
can continuously update the feerate to lower values when the fee estimates
allow it, but never raises the feerate when conditions change in the other
direction.

\begin{figure}
    \centering
    \includegraphics[width=\columnwidth]
    {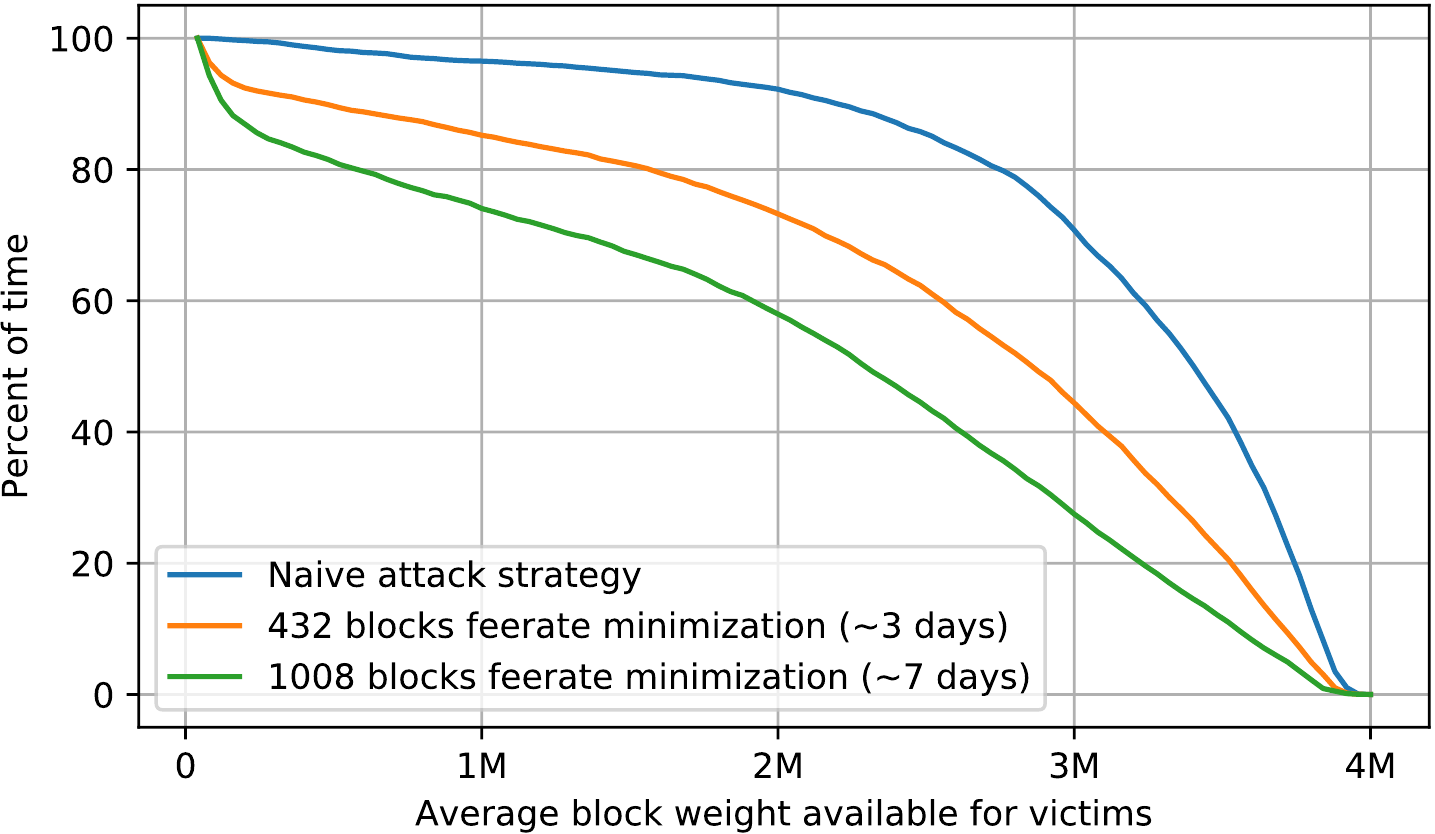}
    \caption{The average block space available for victims' transactions}
    \label{fig:avg-block-space-vs-percent-of-time}
\end{figure}

Figure~\ref{fig:avg-block-space-vs-percent-of-time} shows the effects of using
this feerate minimization strategy. When minimizing the fees over a period of
1008 blocks (approximately 7 days), victims have an average available block
space of 2M (or more) weight units, only 58\% of the time.

\section{Mitigation Techniques} \label{sec:mitigations}
In this section we discuss several mitigation techniques that can help to
reduce the severity of the attack. We start by emphasizing some of the key
points that make the attack feasible.

An important part of a successful attack is the congestion in the underlying
network when the victims close their channels. This of course is not controlled
by the victims, but they can better prepare for it by choosing some parameters
more carefully. These include the time at which the commitment is published,
the maximum allowed number of unresolved HTLCs or the channel's feerate.

Other key points that allow executing a successful attack cannot be controlled
by the victims, and are in fact inherent to the protocol itself. The fact that
the attacker can easily replace victims' transactions after the HTLCs expire,
is based on 3 protocol rules:
\begin{enumerate}[(i)]
    \item The victim can only claim the HTLCs using HTLC-success (local)
          transactions that must be signed by both parties. They therefore
          cannot modify their transactions when faced with an uncooperative
          attacker
    \item The HTLC-success transactions published by the victims are replaceable
          (as we explain in Section~\ref{subsec:RBF})
    \item The attacker's HTLC-timeout transactions do not require the signatures
          of the victims, allowing the attacker to increase the fees
\end{enumerate}

\subsection{Bad Mitigation Strategies}
\smallheadline{Simplifying Incoming HTLCs First}
A possible mitigation that may come to mind (but is ill advised) is to enact
the following change in the HTLC forwarding process: Once the secret preimage
is propagated back from the target, HTLCs in each channel on the path are
resolved. However, instead of having nodes agree to resolve an outgoing HTLC
immediately upon receiving the preimage, one might be tempted to require
that nodes forward the secret back towards the sender, and await for the
corresponding incoming HTLC to be resolved first.

Such a process would seemingly raise difficulties for the attacker. Since the
attacker does not intend to resolve any HTLCs within the first-hop channel, our
current mode of attack will be harder to carry out: No HTLCs will be resolved
on the last hop as well, and the attacker will be forced to publish many HTLC
transactions, and claim them all in a limited amount of time. He will not be
able to increase fees for his transactions, and will in fact find himself in a
similar situation that the victims are in, when attacked. This could make the
attacker lose funds he locked in channels, and thus disincentivize executing an
attack.

The reason we \emph{do not} recommend this change however, is that this
mitigation introduces \emph{a new} potential attack that could spam the
blockchain with many more transactions without additional effort. An attacker
could route many payments from and to his own lightning nodes (similarly to our
attack), using as many intermediate nodes on the payment path. According to
this mitigation, \emph{none} of the HTLCs on the different channels involved in
forwarding the payments could be resolved before they are resolved by the
source node. If the attacker decides not to resolve HTLCs, all channels on the
payment path, including ones the attacker is not part of, will be forced to
close to the blockchain with \emph{all} HTLCs unresolved.

In our attack, each victim's channel generates a number of blockchain
transactions that is proportionate to the number of HTLCs on that channel. In
this new ``spam attack'', this number is multiplied by the length of the
payment path, that could include up to 20 intermediate nodes.

\smallheadline{Increasing HTLC Fees Via Child Pays For Parent}
Another flawed solution is to try and increase the fee paid by the victim's HTLC
transaction using the \emph{Child Pays For Parent} (CPFP) technique. To employ this
method, the victim publishes his HTLC-success transaction (that pays low fees),
denoted $tx_{success}$, along with another transaction, denoted $tx_{cpfp}$. $tx_{cpfp}$ is set up so that it depends on $tx_{success}$ (it spends one of its outputs), and so that it pays a higher fee that supplements that of $tx_{success}$. This incentivizes miners to
include both transactions together.

However, since $tx_{success}$ is replaceable, the rules state that transactions that depend on it are replaceable as well. $tx_{cpfp}$ is therefore replaceable as well, as long as $tx_{success}$ remains unconfirmed. Thus, using CPFP does not prevent the attacker from replacing the transactions and increasing his own
transaction's fee, and would not prevent the theft. The only way to stop the attacker would be to pay fees that are higher than the amount being contested in the HTLC. In this case the attacker gains nothing from the attack, but the victim losses all as well, as the fees effectively burned away all the money that was to be gained.

\subsection{Reducing the Maximum Number of Unresolved HTLCs}
One of the main reasons this attack could work is that there are only so many
HTLC-success transactions that could be confirmed from the moment the
commitments are published until the HTLCs expire. Once an HTLC has expired, the
victim has almost no chance of claiming it. Reducing the maximal number of
unresolved HTLCs allowed at any moment, would lower the number of transactions
that need to enter the blockchain in that time window. An attacker will then
need more victims in order to steal funds from the same amount of HTLCs.
However, finding an optimal value for the maximum allowed HTLCs is not trivial.
As this value decreases, the potential of another attack increases. As shown
in~\cite{mizrahi2020congestion}, a low value for the $max\_accepted\_htlcs$
parameter would help an attacker execute a denial-of-service attack, by
congesting many channels in the network.

\subsection{Deciding When to Close Channels}
Currently, the commitment broadcast delta defines the time (prior to HTLC
expiration) that nodes begin to unilaterally close channels with unresolved
incoming HTLCs.
The commitment broadcast delta should be set high enough to allow the node to
claim these soon-to-expire HTLCs, in case the other party is uncooperative. The
Lightning Network specifications~\cite{bolts} suggest publishing the commitment
7 blocks before expiration.

Default parameters used by different implementations are shown in
Table~\ref{table:commitment-broadcast-delta}, from which we learn that nodes
use a smaller time window (``Commitment broadcast delta'') to claim HTLCs than
they potentially could (``HTLC expiry delta'').

\begin{table}
    \centering
    \begin{tabular}{ |c|c|c|c| }
    \hline
                                               & C-Lightning & LND & Eclair \\
    \hline
    HTLC expiry delta                          & 14          & 40  & 144 \\
    \hline
    \shortstack{Commitment \\ broadcast delta} & 7           & 10  & 6 \\
    \hline
    \shortstack{Maximum \\ accepted HTLCs}     & 30          & 483 & 30 \\
    \hline
    \end{tabular}
    \caption{Default values for different channel's parameters of major
    Lightning implementations}
    \label{table:commitment-broadcast-delta}
\end{table}

Simply increasing the commitment broadcast delta, each attacked node will have
more time (blocks) to claim the HTLCs on the blockchain before they expire and
become spendable by the attacker, making it less probable to lose funds in an
attempted attack.

Such change, however, could lead to premature closure of honest channels in
case the previous node is honest but fails to respond in time. We believe such
changes should not be too damaging, as they are unlikely to happen to honest
nodes. A node that just forwarded an HTLC payment is usually responsive in the
next few seconds, which is oftentimes enough to forward and completely resolve
payments successfully.

A more extreme version of this mitigation is to initiate a closure of the
channel if the previous hop fails to simplify an HTLC within a short period of
time \emph{regardless} of the HTLC's expiration time (this timeout can be
specified in seconds, since blocks may sometimes be created in bursts).

In short, the benefits from closing channels earlier when faced with
misbehaving peers seem to surpass any downside.

\smallheadline{Dynamic closing rules}
If a node has many pending incoming HTLCs that the other party does not
resolve, it could raise the likelihood that the node is being attacked. In that
case, the channel should be closed earlier, to avoid losing funds.

We therefore advocate for a more sophisticated policy that dynamically sets the
commitment broadcast delta for each channel, based on the potential loss that
this channel may incur in case of an attack. E.g. a commitment broadcast delta
of a channel could increase as the number of concurrent unresolved HTLCs
increases or if the total value of HTLCs is high.

\subsection{Do Not Wait for Commitment Tx Confirmation}
When a node decides to unilaterally close one of its channels to claim incoming
HTLCs on the blockchain, it needs to publish its latest commitment transaction
along with a collection of HTLC-success transactions. In practice, some
implementations (e.g., LND and C-Lightning) wait until the commitment
transaction is confirmed before releasing the HTLC-success transactions. This
way, nodes lose precious time that can be used to get HTLC transactions
accepted.

Since all of these transactions can be published to the network immediately, a
simple fix would be to do so. This mitigation technique has no visible
drawbacks and is relatively easy to implement (Eclair nodes already follow this
practice).

\subsection{Reputation-Based Behavior}
As we have shown in our simulations, a high number of maximum allowed HTLCs
helps an attacker to execute a successful attack. However, a high number is
also \emph{desirable} if we want to support greater throughput, and therefore a
high transaction rate.
Similarly, allowing HTLCs that transfer larger amounts also aids the attacker,
but is desirable for the Lightning Network's functionality.

We suggest setting the value of these parameters differently for each channel,
and basing them on a reputation score we assign to the other party with whom we
share the channel. A node will agree to route a greater number of high value
HTLCs together with peers it considers ``reputable''. We thus lower the node's
risk and the potential loss it may suffer in case of an attack.

For example, we can assign a higher reputation score to channels that we
initiate (we chose who the other party is), and give a lower score to channels
that were created based on requests from unknown counterparts. Reputation
scores could grow over time, as a node proves its ability to ``behave well'',
which would imply that attackers would need to run longer-term (and thus more
costly) attacks to have a similar effect.

\subsection{Anchor Outputs}
A new proposal was made recently~\cite{anchor-outputs-github} that provides an
improvement to the way lightning transaction fees are paid.
The proposal introduces another type of commitment output, called \textit{an anchor}
output. Such outputs may be used by each of the nodes to increase the fee of a
commitment, using the \textit{child-pays-for-parent} method, allowing the
commitment to be confirmed sooner. Each party will have its own anchor output in
the commitment, which only it can spend.\footnote{In practice, once the
commitment is deeply confirmed, anchor outputs, whose value is low, will be spendable by anyone, to
prevent a burden on the utxo set}
This proposal also changes the structure of HTLC-claiming transactions
by using {\verb+SIGHASH_SINGLE|SIGHASH_ANYONECANPAY+} signature hash type,
allowing to add inputs and increase the fee. Regarding our attack, this method
would allow the victims to publish their HTLC-success with increased fee,
and therefore better compete with other transactions in the mempool,
which they are not able to do today.

Although the victims would be able to set high fees for their transactions, they
still would not be able to compete with the attacker's transactions once the HTLCs
expire. Since the attacker constructs his transactions as he wishes (without
needing the victims' signatures), he could
make them irreplaceable, preventing the victims from replacing them, regardless of
how much fee they offer.
The attacker's transactions may now need to pay more fees (as the victims could
have increased them), yet, since the expired HTLCs belong to the victims, the
attacker loses nothing by paying them.

\subsection{Non-Replaceable HTLC transactions}
Another possible mitigation would be to prevent the attacker from replacing the
victims' HTLC-success transactions, by making them irreplaceable. It appears that
replaceability is not strictly needed for the trustless nature of the
protocol. Still, this attempt must be done carefully, and care should be
taken not to break the revocation mechanisms.

\subsection{A Recommendation For Exchanges \& Other Payment Recipients}
As we saw in the attack, adding an incoming HTLC to the commitment does not
guarantee the receipt of the payment. The amount of an HTLC payment could be
stolen if it is not resolved and the receiver has to claim it on the blockchain.
Therefore, a payment recipient should not consider a payment as received when it is added
to its channel, but only after the sender resolved it and committed to the new
state (with the HTLC amount moved to the balance of the receiver).
In particular, this applies to any exchange or ATM that accepts deposits through
lightning payments and allows the sender to further transfer or withdraw cash or bitcoins.

\section{Related Work} \label{sec:related-work}
Many aspects of the Lightning Network have been extensively studied. A wide
survey on second-layer solutions was done in~\cite{cryptoeprint:2019:360}.
A technical review of the building blocks of the Lightning Network is presented
in~\cite{mccorry2016towards}.

Some basic attack vectors were mentioned in the original Lightning Network
paper~\cite{poon2016bitcoin}.
One problem that was introduced was a potential blockchain spamming attack,
caused by a malicious party that forces many channels to be closed at the same
time. Yet, no concrete evaluation was done to see how severe such an attack
might be.

The notion of \textit{Channel Exhaustion} is discussed
in~\cite{discharged-payment-channels}. It describes a way to disrupt the normal
operation of a channel as a bi-directional route, by pushing its entire balance
to one side, making it impossible to route payments in one direction.
A detailed channel exhaustion attack is described in~\cite{lockdown-attack}.

A known attack technique called \textit{Griefing}~\cite{griefing-definition} can
be used to execute denial-of-service attacks on the network, by initiating a
multi-hop payment that will be in pending mode for a long period of time,
possibly until it expires entirely. One attack that makes use of griefing is
described in~\cite{mizrahi2020congestion}, where they exploit the HTLC mechanism
to block channels from routing payments in both directions while minimizing the
cost of the attack far lower than a traditional channel exhaustion.

A new proposal has been discussed recently to mitigate griefing 
attacks~\cite{lightning-dev-griefing-attack-mitigation}.
The proposal tries to get an attacker's channels closed if he uses griefing,
which will impose additional costs to the attacker and prevent him from routing
payments and earning fees.
Yet another solution for griefing attacks is proposed
in~\cite{banerjee2020griefing}, where the authors suggest a way to penalize an attacker.

An attack shown in~\cite{malavolta2019anonymous} can
successfully steal payment fees if a malicious party controls two nodes on the 
payment route. This is achieved by excluding all nodes between the two malicious
ones from the normal process of completing an HTLC transaction, and thus earning
all the fees they should have received from forwarding the payment.

Another potential attack known as ``RBF-pinning'' exploits the HTLC mechanism
along with the fact that some Lightning implementations do not monitor the
mempool~\cite{rbf-pinning}. Here, an attacker releases transactions that claim
HTLCs (which include the preimage), but does so using extremely low fees. This
ensures that these transactions do not appear on the blockchain, but rather stay
in mempools, where the lightning implementations do not detect them (and hence
do not learn the published preimages). The transactions published by the
attacker are irreplaceable, so they cannot be replaced by the victim.

Other works study the network from an economic point-of-view.
In~\cite{engelmann2017towards}, the authors explore the relation between the cost of a
lightning payment and the number of hops in its route, and argue that in some
cases, lightning payments can be even more expensive than normal blockchain
transactions. The incentives to participate in the Lightning
Network were analyzed in~\cite{beres2019cryptoeconomic}. According to this work,
participation is in fact \emph{irrational} for most of today's
large nodes.

Payment routing was analyzed
in~\cite{prihodko2016flare,tochner2019hijacking,engelmann2017towards}, and some
improvements were proposed for choosing cheaper routes in a way that is 
also less prone to attacks.

The Lightning Network's topology was studied
in~\cite{seres2020topological,lin2020lightning}. It has been shown that the
network is highly centralized, with a small number of hubs connecting large
parts of the network together. Removing such hubs can partition the network
into many components.

Other works focus on the privacy of participants in the network. The network
aims to keep nodes' balances private, and therefore a channel's exact balance
is never announced. Several methods to deduce the exact balance of a channel
have been suggested~\cite{herrera2019difficulty,cryptoeprint:2019:1385,
tikhomirov2020probing}. The main technique is to request a payment via the
channel, then, based on whether the payment succeeds to traverse the channel or
not, the sender learns whether the balance is greater or smaller than the
requested payment amount.

\section{Conclusions} \label{sec:conclusions}
In this work we presented the \textit{Flood $\&$ Loot} attack, in which an
attacker can successfully steal funds from nodes on the Lightning Network. We
laid concrete steps for how this attack could be executed and evaluated it by
simulating it locally.

We examined the mechanisms that are used by lightning nodes to determine
blockchain fees and studied their influence on the success of the attack. We
also pointed out ways an attacker could use to amplify the attack, based on the
results of the fee estimation mechanisms, that would allow him to steal more
funds for the same effort (initial investment and number of opened channels
required for successful execution).

We continued with a discussion of multiple mitigation techniques that could
reduce the scope of the attack, by taking into account the fundamental elements
that contributed to its success. Still we believe that in many ways the
vulnerabilities exploited by our attack are inherent to the way HTLCs work, and
thus the attack cannot be avoided completely.

\begin{acks}
    We would like to thank Bastien Teinturier for his valuable feedback.
    This research was supported by the Israel Science Foundation (grant 1504/17)
    and by a grant from the HUJI Cyber Security Research Center in conjunction
    with the Israel National Cyber Bureau.
\end{acks}

\bibliographystyle{paper/ACM-Reference-Format}
\bibliography{references}

\end{document}